\documentstyle[pra,aps,epsf]{revtex}

\begin{document}

\twocolumn[\hsize\textwidth\columnwidth\hsize\csname
@twocolumnfalse\endcsname

\title{Generation of polarization-entangled photon pairs in a cascade of\\
two type-I crystals pumped by femtosecond pulses}
\author{Yoshihiro Nambu,${^{1}}$ Koji Usami,${^{2}}$ Yoshiyuki Tsuda,$^{3,4}$ Keiji
Matsumoto,$^{4}$ and Kazuo Nakamura${^{1,2}}$}
\address{${{}^1}$Fundamental Research Laboratories, NEC Corporation, 
34, Miyukigaoka, Tsukuba, Ibaraki 305-8501, Japan}
\address{${{}^2}$Depertment of Material Science and Engineering, Tokyo
Institute of Technology, \\
4259, Nagatsuta-chou, Midori-ku, Yokohama, Kanagawa, 226-0026, Japan}
\address{$^{3}$Institute of Mathematics, University of Tsukuba, Tsukuba, Ibaraki, 305-8571, Japan}
\address{$^{4}$ERATO, Japan Science and Technology Corporation \rm(JST),\it 5-28-3, Hongo, Bunkyo-ku, Tokyo, 113-0033, Japan}
\date{\today}
\maketitle

\begin{abstract}
We report the generation of polarization-entangled photons by
femtosecond-pulse-pumped spontaneous parametric down-conversion in a cascade
of two type-I crystals. Highly entangled pulsed states were 
obtained by introducing a temporal delay between the two orthogonal polarization
components of the pump field. They exhibited high-visibility quantum
interference and a large concurrence value, without the need of post-selection
using narrow-bandwidth-spectral filters. The results are well explained
by the theory which incorporates the space-time dependence of interfering
two-photon amplitudes if dispersion and birefringence in the crystals are
appropriately taken into account. Such a pulsed entangled photon well localized in time
domain is useful for various quantum communication experiments,
such as quantum cryptography and quantum teleportation.
\end{abstract}


\vskip1pc]

\narrowtext

\section{INTRODUCTION}

\label{SEC1}

Photon pairs generated in the process of spontaneous parametric
down-conversion (SPDC) have been an effective and convenient source of
two-particle entangled states used in tests on the foundations of quantum
mechanics as well as application to quantum information technologies such as
quantum cryptography and quantum teleportation. In such applications, a
pulsed source of entangled photon pairs is particularly useful because the
times of emission are known to the users. For example, in the
entanglement-based quantum key distribution protocol\cite%
{Ekert92,Jennewein00,Naik00}, it is convenient to share entangled photons
between sender and receiver in a train of short optical pulses synchronized
with the common timing clock pulses. Such a pulsed source allows the sender
and receiver to know the arrival times of the photons within the duration of
the pump pulse and to time stamp each received bit readily so that they can
sift their raw data to generate a shared key from their public discussion. A
pulsed source is also crucial in a certain class of experiments that require
the use of several photon pairs at a time, such as quantum teleportation\cite%
{Bennett93,Bouwmeester97,Bouwmeester98}, entanglement distillation\cite%
{Bennett96}, and the generation of multiphoton entangled states\cite%
{Greenberger90,Bouwmeester99,Pan00}. In these experiments, time-synchronized
entangled photon pairs must be available. It has also been pointed out that
primitive elements of a quantum computer can be constructed from the
combined use of several entangled photons and Bell-state measurement\cite%
{Gottesman99,Gottesman99-2}.

A great deal of effort has been devoted to developing a pulsed source of
entangled photon pairs. Over the past decade, femtosecond-pulse-pumped SPDC
has been extensively studied by several groups. It has been shown that the
entangled states generated in type-II SPDC pumped by ultrashort pulses shows
disappointingly low visibility of the quantum interference, and narrow-
bandwidth filters are required to increase the visibility at the expense of
the photon flux\cite{Guiseppe97,Grice97,Grice98}. Theoretical studies
revealed that the down-converted photons are entangled simultaneously in
polarization and space-time, or equivalently, wave number-frequency because
of the significant effects of dispersion and birefringence in the crystal%
\cite{Guiseppe97,Keller97,Kim01}. As a result of this multiple-entanglement,
it is possible to distinguish, with some degree of certainty, which of the
interfering pathways occurs by the measurement that distinguishes the
space-time component of the state, for example, the measurement of the
arrival time of the photons at the detector. This distinguishing space-time
information is enough to seriously degrade the visibility of two-photon
polarization interference. The presence of this distinguishing space-time
information stems from the fact that the down-converted photons are emitted
spontaneously, but the times of emission are known to be within the duration
of the pump pulse. The resultant two-photon state is well localized in
space-time, and provide undesirable timing capability. The situation is
serious in the ultrashort-pulse-pumped SPDC, but is absent in the ordinary
cw-pumped SPDC. Therefore, it is vital in the ultrashort-pulse-pumped SPDC
to ensure that the space-time component of the state contained no
distinguishing ``which-path'' information for the photons in order to
observe the entanglement and quantum interference in the polarization degree
of freedom. Various interferometric techniques have been proposed to
eliminate the entanglement in unnecessary degrees of freedom and to recover
the quantum interference\cite{Kim01,Kim2000,Kim01a,Branning99,Branning00}.

On the other hand, cw-pumped SPDC exhibits a polarization-entanglement with
high-visibility quantum interference capability. In particular, Kwiat et al.
realized maximally polarization-entangled photons using noncollinear SPDC
with type-II phase matching\cite{Kwiat95}. More recently, they demonstrated
that noncollinear SPDC using two spatially separate type-I nonlinear
crystals pumped by a cw laser exhibits high-visibility quantum interference%
\cite{Kwiat99}. The latter source is particularly convenient since the
desired polarization-entangled states are produced directly out of the
nonlinear crystal with unprecedented brightness and stability and without
the need for critical optical alignment. Their result attracted great
attention for the experimentalists. Recently, Kim et al. showed that
polarization-entangled photons from two spatially separated type-I nonlinear
crystals pumped by femtosecond laser pulses exhibit high-visibility
interference\cite{Kim2000}. In their experiment, collinear degenerate type-I
SPDC and a beam splitter were used to create a polarization-entangled state.
The state prepared after the beam splitter was, however, not considered to
be entangled without amplitude postselection. Only when one considers those
postselected events (half of the total events) in which photons traveled to
different output ports, one can observe quantum interference. Later, this
problem was solved by using collinear nondegenerate type-I SPDC and a
dichroic beam splitter\cite{Kim01a}. It was shown that mostly maximally
entangled states showing high-visibility (92\%) quantum interference were
obtained without amplitude and spectral postselection just before detection.
However, the available flux of the entangled photon pairs was still very
limited ($\leq $100 s$^{-1}$).

In this paper, we apply femtosecond-pulse pumping to the second Kwiat scheme
with type-I phase-matching arrangement. We shall demonstrate that highly
entangled polarization states with sufficiently large flux are obtained when
an appropriate temporal compensation is given in the pump pulses. The
results will be shown to be consistent with the theory which incorporates
the space-time dependence of interfering two-photon amplitudes, if the
effects of dispersion and birefringence in the two-crystals are
appropriately taken into account.

\section{EXPERIMENT}

\label{SEC2}

\begin{figure}[tbp]
\epsfxsize=9.0cm
\centerline{\epsfbox{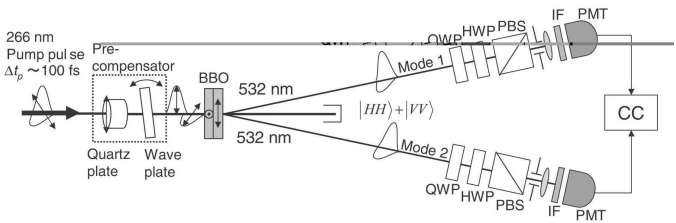}}
\caption{Experimental setup for generating and characterizing polarization
entangled pulsed photons. The pump pulse polarized at 45${{}^{\circ }}$
irradiates a cascade of two type-I crystals, whose optic axes are
orthogonally oriented. Pre-compensator (quartz plate and tiltable wave
plate) introduces relative delays between the two-photon amplitudes created
in the first and second crystal. CC is the coincidence circuit.}
\label{F1}
\end{figure}

Our experimental setup is schematically shown in Fig. \ref{F1}. Two
adjacent, thin, type-I crystals (BBO) whose optic axes are horizontally (%
{\it H}) and vertically ({\it V}) oriented, respectively, are pumped by 45${%
{}^{\circ }}$- polarized femtosecond pulses at 266 nm. The pump pulses were
third harmonic of a mode-locked output of Ti/Sapphire laser, whose
approximate average power was 150 mW and repetition rate was 82 MHz. Due to
type-I coupling, {\it H}-polarized photon pairs at 532 nm are generated by
the {\it V}-polarization component of the pump field in the first crystals,
and {\it V}-polarized photon pairs are generated by the {\it H}-polarization
component of the pump field in the second crystal. These two possible
down-conversion processes equally likely occurs and are coherent with one
another\cite{Ou89,Ou90}. The SPDC was performed under a degenerate and
quasi-collinear condition. The signal and idler photons making an angle of 3$%
{{}^{\circ }}$ with respect to the pump laser beam and having the same
wavelength around 532 nm were observed through the identical interference
filters, centered at 532 nm, placed in front of the detectors. The
polarization correlations were measured using polarization analyzers, each
consisting of a rotatable half- wave plate (HWP) and a quarter-wave plate
(QWP) (for 532 nm) followed by a polarizing beam splitter (PBS). After
passing through adjustable irises, the photons were collected using
60-mm-focal-length lenses, and directed onto the detectors. The detectors
(PMT) were photomultipliers (HAMAMATSU H7421-40) placed at $\sim $1.5 m from
the crystal, with efficiencies of $\sim $40\% at 532 nm and dark count rates
of the order of 80 s$^{-1}$. The photodetection area was about 5 mm in
diameter. The outputs of the detectors were recorded using a time interval
analyzer (YOKOGAWA TA-520), and pulse pairs received within a time window of
7 ns were counted as coincident.

\begin{figure}[tbp]
\epsfxsize=9.0cm
\centerline{\epsfbox{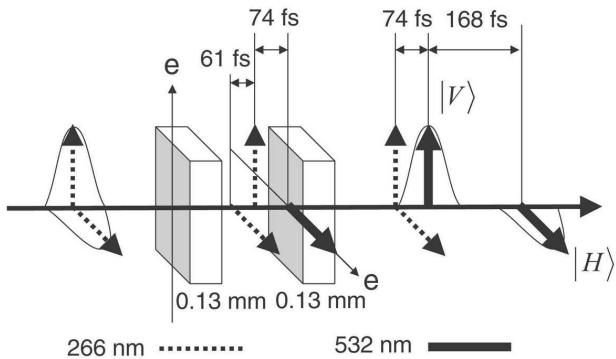}}
\caption{Schematic of the temporal development of the down-converted photons
and the pump. See text for details.}
\label{F2}
\end{figure}

To obtain a truly polarization-entangled state, care must be taken to
disentangle the polarization degree of freedom from any other degrees of
freedom, that is, to factorize the total state into product of the
polarization- entangled state and those describing other degrees of freedom.
This is equivalent to saying that effective polarization entanglement
requires the suppression of any distinguishing information in the other
degrees of freedom that can provide potential information about
``which-polarization'' the emitted pair have. In our case, since which
crystal the origin of each pair is and their polarization are intrinsically
correlated, distinguishing ``which-crystal'' information must be also
eliminated. Accordingly, to make emitted spatial modes for a given pair
indistinguishable for the two crystals, we overlap the down-conversion light
cones spatially by using very thin crystals each having $\sim $130 $\mu $%
m-thickness. In addition, it is important to eliminate distinguishing
space-time information inherent in the two-photon states produced in the
ultrashort-pulse-pumped SPDC. Figure \ref{F2} schematically illustrates what
may happen when a 45${{}^{\circ }}$- polarized femtosecond pulse incidents
on two cascaded type-I BBO crystals, which is estimated roughly from the
optical characteristics of the BBO\cite{Eimerl87}. For simplicity, we
consider here the degenerate collinear SPDC. In the first crystal, {\it H}%
-polarized down-converted photon wavepackets are advanced at $\sim $74 fs
relative to the {\it V}-polarized pump pulse due to group velocity
dispersion, while the {\it H}-polarized pump pulse is delayed at $\sim $61
fs relative to the {\it V}-polarized pump pulse due to birefringence in the
BBO. In the second crystal, {\it V}-polarized down-converted photon
wavepackets are advanced at $\sim $74 fs relative to the {\it H}-polarized
pump pulse due to group velocity dispersion, while the {\it H}-polarized
down-converted photon wavepackets are advanced at $\sim $242 fs relative to
the {\it H}-polarized pump pulse due to an advance given in the first
crystal ($\sim $135 fs) and dispersion in the second crystal ($\sim $107
fs). Consequently, after the crystals, space-time components of the
two-photon state associated with the polarization states $\left|
HH\right\rangle $ and $\left| VV\right\rangle $ are expected to be
temporally displaced by $\sim $168 fs, where $\left| H\right\rangle $ ($%
\left| V\right\rangle $) means a single photon linearly polarized along a
horizontal (vertical) axis and the first (second) letter corresponds to the
signal (idler). Note that the distance between the two BBO crystals is not
an important factor in this consideration. Since down-converted photon
wavepackets are expected to have widths comparable to that of pump field ($%
\sim $150 fs), it is suggested that the space-time components associated
with $\left| HH\right\rangle $ and $\left| VV\right\rangle $ do not overlap
in space-time. As a result, ``which-polarization'' information may be
available, in principle, from the arrival time of the photons at the
detector.

To eliminate this distinguishing space-time information, a polarization
dependent optical delay line for the 266-nm pump was inserted before the
crystals, which is denoted as the pre-compensator in Fig. \ref{F1}. It
consists of quartz plates, whose optic axes are oriented either vertically
or horizontally, and a Bereck-type tiltable polarization compensator. The
combination of the quartz plates of various thicknesses provides a relative
delay of $\left| T\right| \leq $350 fs between the {\it H}- and {\it V}%
-polarization components of the 266-nm pump field, while the sign is given
by the orientation of its optic axis. This delay line can compensate for
relative delay between the state associated with $\left| HH\right\rangle $
created in the first crystal relative to the state associated with $\left|
VV\right\rangle $ created in the second crystal to overlap them temporally.
The Bereck-type polarization compensator is used to adjust the subwavelength
delay. As a result, a two-photon Bell state $\left| \Phi ^{+}\right\rangle
=(\left| HH\right\rangle +\left| VV\right\rangle )/\sqrt{2}$ can be directly
created.

Two kinds of experiments were carried out to confirm whether the target
state $\left| \Phi ^{+}\right\rangle $ was successfully prepared. The first
experiment was a quantum state tomography\cite{White99}. The polarization
density matrices were estimated from 16 kinds of joint projection
measurements performed on an ensemble of identically prepared photon pairs.
These joint measurements consist of four kinds of projection measurements
onto \{$\left| H\right\rangle ,\left| V\right\rangle ,\left| D\right\rangle
,\left| L\right\rangle $\} on each member of a photon pair, where $\left|
D\right\rangle =(\left| H\right\rangle +\left| V\right\rangle )/\sqrt{2}$
and $\left| L\right\rangle =(\left| H\right\rangle +i\left| V\right\rangle )/%
\sqrt{2}$. We used a maximum likelihood calculation\cite{James01} to
estimate the density matrix. We also calculated the concurrence of the
state, which is known to give a good measure for the entanglement of a
two-qubit system\cite{Wootters98}. The second experiment was a conventional
two-photon polarization interference experiment. In this experiment, QWPs in
mode 1 and mode 2 were set to 0${{}^{\circ }}$, i.e., their optic axes were
vertically oriented, while HWP in mode 1 was set to $\pm $22.5${{}^{\circ }}$
and HWP in mode 2 was rotated. In our setup, this corresponds to the
projection measurements onto \{$\left| L\right\rangle $ or $\left|
R\right\rangle $\}$\otimes \left| \theta \right\rangle $, where $\left|
R\right\rangle =(\left| H\right\rangle -i\left| V\right\rangle )/\sqrt{2}$
and $\left| \theta \right\rangle =\cos 2\theta \left| H\right\rangle +i\sin
2\theta \left| V\right\rangle $. The visibility of an observed interference
pattern gives another convenient measure for the entanglement.

\begin{figure}[tbp]
\epsfxsize=16.0cm
\centerline{\epsfbox{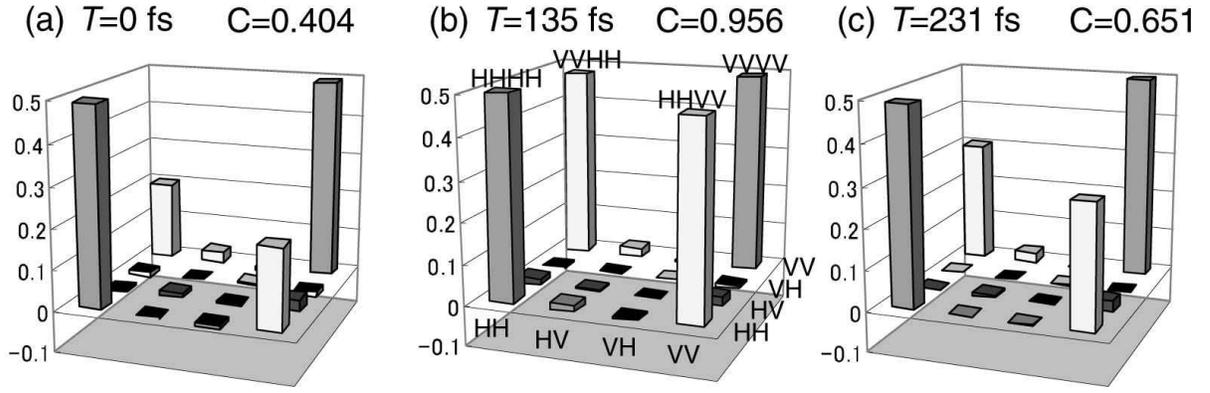}}
\caption{Estimated polarization density matrices for (a) no relative delay,
(b) relative delay of 135 fs, and (c) relative delay of 231 fs between the 
{\it H}- and {\it V}-polarization components of the pump field. Only the
real parts of the matrices are shown. The contribution of the imaginary
parts is negligible. The dark-colored bars show diagonal elements, while
the light-colored bars show off-diagonal elements. Estimated values of
concurrence are shown as $C$.}
\label{F3}
\end{figure}

\begin{figure}[tbp]
\epsfxsize=16.0cm
\centerline{\epsfbox{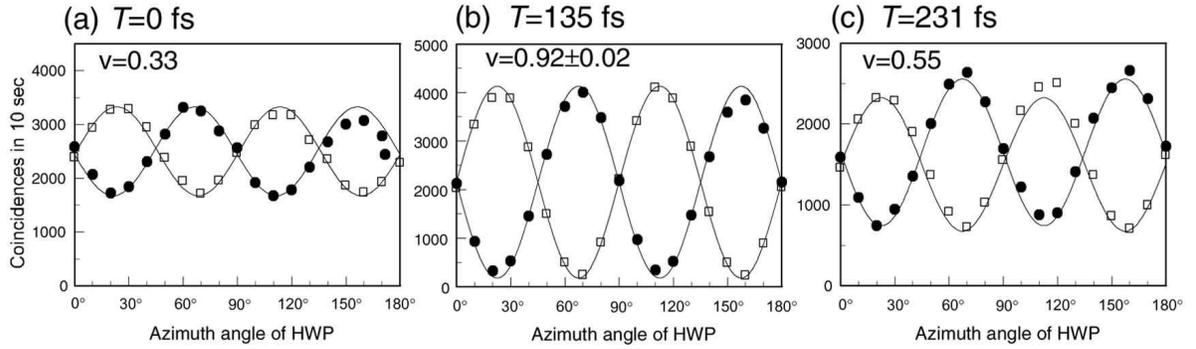}}
\caption{ Polarization correlation experiment. Coincidence counts in 10 s is
plotted against the orientation angle of the half-wave plate in mode 2,
while the half-wave plate in mode 1 was fixed to $\pm 25.5{{}^\circ}$, for
the states shown in (a)-(c) of Fig. \ref{F3}. Estimated visibilities are
also shown in the figure.}
\label{F4}
\end{figure}

\begin{figure}[tbp]
\epsfxsize=9.0cm
\centerline{\epsfbox{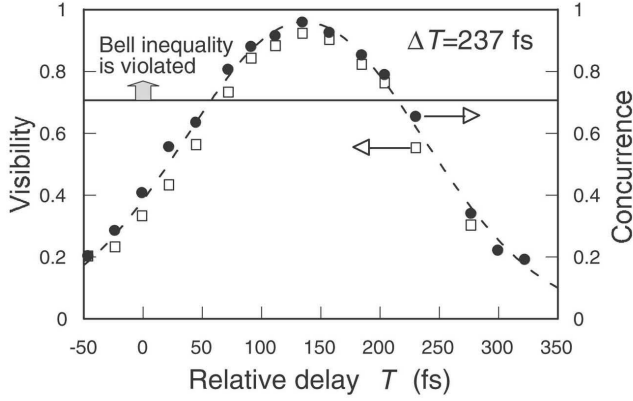}}
\caption{Evaluated concurrence {\it C} and visibility {\it v} when 8-nm
interference filters were used{\it . }These values are plotted against the
relative delay introduced between two orthogonal components of the pump
field. The broken line is Gaussian fit to the data with 237-fs widths
(FWHM). }
\label{F5}
\end{figure}

At first, we used the interference filters having a full-width half maximum
(FWHM) bandwidth of 8 nm in front of the detectors. Figure \ref{F3}
illustrates the estimated polarization density matrices prepared when (a) no
relative delay, (b) relative delay of 135 fs, and (c) relative delay of 231
fs were introduced between the {\it H}- and {\it V}-polarization components
of a 266-nm pump field. In these experiments, maximum coincidence counts and
typical accidental coincidence counts were $\sim $450 cps and \mbox{$<$}10
cps, respectively. As shown by these examples, we could successfully prepare
the states approximately described by $\rho (v)=\frac{1-v}{2}(\left|
HH\right\rangle \left\langle HH\right| +\left| VV\right\rangle \left\langle
VV\right| )+v\left| \Phi ^{+}\right\rangle \left\langle \Phi ^{+}\right| $.
These states should exhibit interference in coincidence rates for the two
detectors in proportion to $P(\theta )=\frac{1}{4}(1\pm v\sin 4\theta )$.
Figure \ref{F4} shows the coincidence rates as a function of the HWP angle
in mode 2 and the evaluated visibilities associated with the states in Fig. %
\ref{F3}(a)-(c). Note that, no background was subtracted in these results,
so that the evaluated visibilities give the lower limits of $v$. The
experiments were repeated for various values of delay $T$ introduced in the 
{\it H}- and {\it V}-polarization components of the pump field. The
evaluated concurrence {\it C} and visibility {\it v} of the polarization
interference are plotted in Fig. \ref{F5} as a function of relative delay $T 
$. This figure clearly indicates that there is a strong correlation between
the concurrence and the visibility. This is quite natural because $C=v$
should hold for the states $\rho (v)$ given by the above form. In this
experiment, the maximum values of the concurrence and visibility were 0.95
and 0.92$\pm $0.02, respectively. The broken line in Fig. \ref{F5} is a
Gaussian curve fitted to the measured data, and the FWHM width is 237 fs.

Next, to investigate the effects of interference filters used in front of
the detectors, we replaced them with those having a FWHM bandwidth of 40 nm.
Figure \ref{F6} shows {\it C} and {\it v} evaluated by the same procedure,
except that the backgrounds have been subtracted for both the concurrence
and the visibility data in this case. The maximum values of {\it C} and {\it %
v} were slightly reduced to 0.92 and 0.90, respectively, while the maximum
coincidence counts were increased by a factor of $\approx $ 6.
(Unfortunately, the accidental coincidence increased more than six times,
which is probably due to ambient light.) The broken line in Fig. \ref{F6} is
a Gaussian curve fitted to the measured data, and the FWHM width is $\approx 
$205 fs. The width of the fitted curve was slightly reduced in comparison
with that for the 8-nm filters, which was due to the spectral filtering
effects, as shown later.

We also measured the autocorrelation of the pump field by measuring the
degree of polarization, or equivalently, the magnitude of the Stokes
parameter $\left| {\bf s}\right| =\sqrt{s_{1}^{2}+s_{2}^{2}+s_{3}^{2}}$ of
the pump just after the pre-compensator (and before the crystals), where $%
s_{1}$, $s_{2}$, and $s_{3}$ are three independent Stokes parameters. Simple
analysis shows that $\left| {\bf s}\right| $ is proportional to the temporal
overlap of the {\it H}- and {\it V}- polarization components of the pump
field between which relative delay $T$ was introduced by the
pre-compensator, and directly gives the autocorrelation of the pump field.
Figure \ref{F7} shows the measured $\left| {\bf s}\right| $ as a function of 
$T$. We found that $\left| {\bf s}\right| $ can be fitted with a Gaussian
curve having a FWHM width $\approx $216 fs. Assuming the transform limited
Gaussian-shape pulse, the FWHM temporal width of the pump field is estimated
to be $\approx $153 fs, and that of the pump intensity to be $\approx $108
fs. We see that the width of the autocorrelation of the pump field is
comparable to that of the measured concurrence and visibility data shown in
Figs. \ref{F5} and \ref{F6}. This result suggests that the width of the
measured concurrence and visibility data are determined chiefly by the
temporal width of the pump field.

\section{THEORETICAL ANALYSIS}

\label{SEC3}

\begin{figure}[tbp]
\epsfxsize=9.0cm
\centerline{\epsfbox{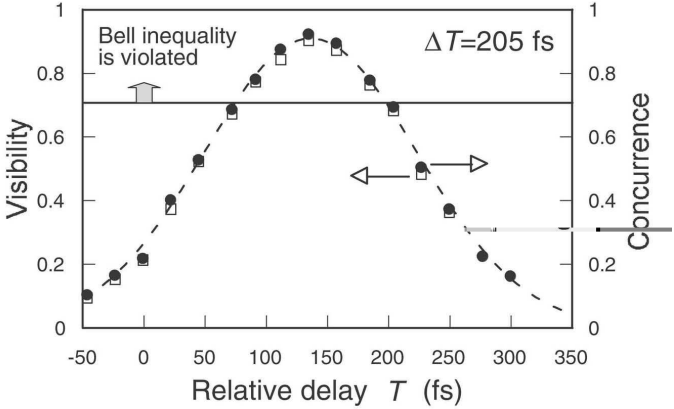}}
\caption{Evaluated concurrence {\it C} and visibility {\it v} when 40-nm
interference filters were used{\it .} The broken line is Gaussian fit to the
data with 205-fs widths (FWHM).}
\label{F6}
\end{figure}

Experimental results are analyzed based on the theory given by Grice et al.%
\cite{Grice97}, Keller et al.\cite{Keller97}, and Kim et al.\cite{Kim01}. In
the following analysis, we will focus our attention on the Gaussian curve
fitted with the measured concurrence and visibility data shown in Figs. \ref%
{F5} and \ref{F6}, which will be denoted by $V(T)$. In particular, we will
analyze how the widths of these curves are determined and depend on the
experimental parameters. To make the following discussion self-contained, it
may be helpful to begin by showing the minimum of the previous theories
before discussing our problem. Let us consider the degenerate,
quasi-collinear SPDC in a {\it single} type-I crystal pumped by a field with
the central frequency $\Omega _{p}$ and the FWHM temporal width $\tau
_{p}^{FWHM}$. For simplicity we ignore the transverse components of the
photon wave vector. Let $\omega _{p}[\omega _{s},\omega _{i}]$ be the
frequency and $k_{p}(\omega _{p})[k_{s}(\omega _{s}),k_{i}(\omega _{i})]$
the wave number for the pump [signal, idler]. The perfect phase-matching
condition is given by $\Omega _{s}=\Omega _{i}=\Omega _{p}/2$ for the
frequency and $K_{o}(\Omega _{s})+K_{o}(\Omega _{i})=K_{e}(\Omega _{p})$ for
the wave number, where $K_{j}(\Omega )$ denotes the wave number for a photon
having a frequency $\Omega $ and polarization along the ordinary ({\it j}=%
{\it o}) or extraordinary ({\it j}={\it e}) optic axis. To simplify the
notation, we introduce detuning for the signal, the idler, and the pump from
the perfect phase-matching condition by $\nu _{s}=$ $\omega _{s}-\Omega
_{p}/2$, $\nu _{i}=$ $\omega _{i}-\Omega _{p}/2$, and $\nu _{p}=\omega
_{p}-\Omega _{p}$, respectively. To first-order in the interaction, the
general expression of the two-photon state after the crystal is\cite{Grice97}
\begin{equation}
\left| \Psi \right\rangle =\frac{1}{2\pi }\int \int d\nu _{s}d\nu _{i}\Psi
(\nu _{s},\nu _{i})a^{\dagger }(\nu _{s})b^{\dagger }(\nu _{i})\left|
0\right\rangle ,  \label{eq1}
\end{equation}%
where $a^{\dagger }(\nu _{s})$ and $b^{\dagger }(\nu _{i})$ are the photon
creation operators for the o-polarized signal and idler modes characterized
by detuning frequencies $\nu _{s}$ and $\nu _{i}$, respectively, which are
defined after the crystal. The function $\Psi (\nu _{s},\nu _{i})$ is the
two-photon amplitude in the frequency domain, and is given by \cite{Grice97} 
\begin{equation}
\Psi (\nu _{s},\nu _{i})=\left\langle 0\right| a(\nu _{s})b(\nu _{i})\left|
\Psi \right\rangle =C\alpha (\nu _{s}+\nu _{i})\Theta (\nu _{s},\nu _{i}).
\label{eq2}
\end{equation}%
In this expression, $\alpha (\nu _{p})$ is the spectral envelop function of
the time-dependent classical pump field at the input surface of the
crystals, $\Theta (\nu _{s},\nu _{i})$ is the longitudinal phase-matching
function\cite{Grice97}, and {\it C} is a constant. Assuming a Gaussian shape
for the temporal envelop function of the pump field, $\alpha (\nu _{p})$ is
given by 
\begin{equation}
\alpha (\nu _{p})=e^{-\left( \frac{\nu _{p}}{\sigma _{p}}\right) ^{2}},
\label{eq3}
\end{equation}%
where $\sigma _{p}=\sigma _{p}^{FWHM}/2\sqrt{\ln 2}$ $=4\sqrt{\ln 2}/\tau
_{p}^{FWHM}$ is the FWHM bandwidth of the pump field. If z-direction is
taken to be the pump direction, the longitudinal phase-matching function $%
\Theta (\nu _{s},\nu _{i})$ is given by 
\begin{equation}
\Theta (\nu _{s},\nu _{i})=\frac{1}{L}\int_{-L/2}^{L/2}dze^{-i\Delta (\nu
_{s},\nu _{i})z}=\mathop{\rm sinc}[\frac{\Delta (\nu _{s},\nu _{i})}{2}L],
\label{eq4}
\end{equation}%
where {\it L} is the crystal length, $\Delta (\nu _{s},\nu _{i})=k_{s}(\nu
_{s}+\Omega _{p}/2)+k_{i}(\nu _{i}+\Omega _{p}/2)-k_{p}(\nu _{p}+\Omega
_{p}) $ is the phase mismatch and approximately given by\cite{Keller97,Kim01}%
\begin{equation}
\Delta (\nu _{s},\nu _{i})=D_{+}\left( \nu _{s}+\nu _{i}\right) +\frac{1}{4}%
D^{\prime \prime }\left( \nu _{s}-\nu _{i}.\right) ^{2}.  \label{eq5}
\end{equation}%
In this expression, $D_{+}$ and $D^{\prime \prime }$ are crystal parameters: 
$D_{+}\equiv 1/u_{o}(\Omega _{p}/2)-1/u_{e}(\Omega _{p})$ is the group
velocity mismatch where $u_{o}(\Omega _{p}/2)$ ($u_{e}(\Omega _{p})$) is the
group velocity of the o-polarized down-converted photons (the e-polarized
pump photon) inside the crystal and $D^{\prime \prime }\equiv
d^{2}K_{o}/d\Omega ^{2}|_{\Omega =\Omega _{p}/2}$ is the group velocity
dispersion for the down-converted photons.

\begin{figure}[tbp]
\epsfxsize=9.0cm
\centerline{\epsfbox{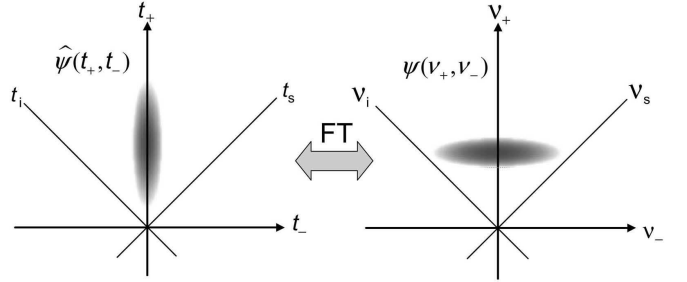}}
\caption{Illustration of two-photon amplitudes in the time and frequency
domains generated by the femtosecond-pulse pump. For simplicity the case
where $L_{s}=L_{i}=L$ is dipicted. They are connected by a Fourier transform
relation.}
\label{F8}
\end{figure}

Since we are interested in the temporal characteristics of the
down-converted photons, we next consider the two-photon state after the
crystal in the time domain. Formally, it can be written as 
\begin{equation}
\left| \Psi \right\rangle =\int \int dt_{s}dt_{i}\hat{\Psi}%
(t_{s},t_{i})a^{\dagger }(t_{s})b^{\dagger }(t_{i})\left| 0\right\rangle ,
\label{eq6}
\end{equation}%
where $a^{\dagger }(t_{s})$ and $b^{\dagger }(t_{i})$ are Fourier transforms
of $a^{\dagger }(\nu _{s})$ and $b^{\dagger }(\nu _{i})$. For example, 
\begin{equation}
a^{\dagger }(t_{s})=\frac{1}{\sqrt{2\pi }}\int d\nu _{s}e^{i\left( \nu
_{s}+\Omega _{p}/2\right) (t_{s}-L_{s}/c)}a^{\dagger }(\nu _{s}),
\label{eq7}
\end{equation}%
where $L_{s}$ is the optical path length from the output face of the crystal
to the observation point. Physically, $t_{s}$ and $t_{i}$ are interpreted as
the time of arrival of the idler and signal at the points that are $L_{s}$
and $L_{i}$ away from the output face of the crystal\cite{Rubin94}, and $%
\hat{\Psi}(t_{s},t_{i})$ gives information about the space-time motion of
the two-photon wavepackets. The two-photon amplitude in the time domain is
simply given by the Fourier transform of the two-photon amplitude in the
frequency domain and by\cite{Rubin94}

\begin{eqnarray}
\hat{\Psi}(t_{s},t_{i}) &=&\left\langle 0\right| a(t_{s})b(t_{i})\left| \Psi
\right\rangle  \nonumber \\
&=&\frac{e^{-i\Omega _{p}t_{s}}}{2\pi }\int \int d\nu _{s}d\nu _{i}e^{-i(\nu
_{s}t_{s}+\nu _{i}t_{i})}\Psi (\nu _{s},\nu _{i}).  \label{eq8}
\end{eqnarray}%
We can further change the variables according to $\nu _{+}=$ $\nu _{s}+\nu
_{i}=\nu _{p}$, $\nu _{-}=\nu _{s}-\nu _{i}$, $t_{+}=(t_{s}+t_{i})/2$, and $%
t_{-}=(t_{s}-t_{i})/2$, and obtain the equivalent Fourier transform relation

\begin{eqnarray}
\hat{\psi}(t_{+},t_{-}) &=&\frac{e^{-i\Omega _{p}t_{s}}}{2\pi }  \nonumber \\
&&\times \int \int d\nu _{+}d\nu _{-}e^{-i(\nu _{+}t_{+}+\nu _{-}t_{-})}\psi
(\nu _{+},\nu _{-}),  \label{eq9}
\end{eqnarray}%
where $\psi (\nu _{+},\nu _{-})=\Psi (\frac{\nu _{+}+\nu _{-}}{2},\frac{\nu
_{+}-\nu _{-}}{2})$ and $\hat{\psi}(t_{+},t_{-})=\hat{\Psi}%
(t_{+}+t_{-},t_{+}-t_{-})$. Physically, $t_{-}$\ and $t_{+}$ mean half the
difference in the arrival time, and the mean arrival time of the idler and
signal photons, at the points that are $L_{s}$ and $L_{i}$ away from the
output face of the crystal, respectively. The absolute square of the
two-photon amplitude $\left| \hat{\psi}(t_{+},t_{-})\right| ^{2}$ is
proportional to the probability of arrival of the idler and signal at the
given points for the given mean arrival time $t_{+}$ and half the difference
in the arrival time $t_{-}$. We show schematically the relevant amplitudes
in the time domain and the frequency domain in Fig. \ref{F8}.

Now, let us proceed to provide a phenomenological model that is consistent
with our experiment using two crystals. We assume down-conversion processes
are equally likely to occur in two crystals whose optic axes are
orthogonally oriented and thy are coherent with one another. We represent
the overall two-photon state after the crystals as a superposition of the
states arising from the SPDC in each crystal with appropriate phase. Then
the two-photon state after the crystals in the time domain can be written as 
\begin{eqnarray}
\left| \Psi \right\rangle &=&\frac{1}{\sqrt{2}}\int \int dt_{+}dt_{-} 
\nonumber \\
&&\times \{\hat{\psi}_{HH}(t_{+},t_{-})a_{H}^{\dagger
}(t_{+}+t_{-})b_{H}^{\dagger }(t_{+}-t_{-})  \nonumber \\
&&+e^{i\phi }\hat{\psi}_{VV}(t_{+},t_{-})a_{V}^{\dagger
}(t_{+}+t_{-})b_{V}^{\dagger }(t_{+}-t_{-})\}\left| 0\right\rangle  \nonumber
\\
&=&\frac{1}{\sqrt{2}}\int \int dt_{+}dt_{-}  \nonumber \\
&&\times \{\hat{\psi}_{HH}(t_{+},t_{-})\left| H,t_{+}+t_{-}\right\rangle
_{s}\left| H,t_{+}-t_{-}\right\rangle _{i}  \nonumber \\
&&+e^{i\phi }\hat{\psi}_{VV}(t_{+},t_{-})\left| V,t_{+}+t_{-}\right\rangle
_{s}\left| V,t_{+}-t_{-}\right\rangle _{i}\}  \label{eq10}
\end{eqnarray}%
where the first (second) term represents the two-photon state generated in
the first (second) crystal, and $\phi $ is the phase difference between the
states generated in the two crystals. In Eq. \ref{eq10} we incorporated a
new dichotomic variable in the creation operators and the two-photon
amplitudes that specifies the polarization ({\it H} or {\it V}) of the
photon, and wrote $\left| H,t_{+}+t_{-}\right\rangle _{s}=a_{H}^{\dagger
}(t_{+}+t_{-})\left| 0\right\rangle $ etc., which denotes a single-photon
state for the signal (or idler) mode characterized by a definite
polarization and time. From the previous discussion, two two-photon
amplitudes $\hat{\psi}_{HH}(t_{+},t_{-})$ and $\hat{\psi}_{VV}(t_{+},t_{-})$
are expected to be temporally displaced by $\tau $ in the $t_{+}$-direction
due to dispersion and birefrengence in the crystals. On the other hand, the
pre-compensator for the pump before the crystals introduces temporal
displacement in the $t_{+}$-direction for these amplitudes by $-T_{p}$.
Therefore, we postulate that these two amplitudes can be written using a
single common amplitude as $\hat{\psi}_{HH}(t_{+},t_{-})=\hat{\psi}%
_{VV}(t_{+}+T,t_{-})\equiv \hat{\psi}(t_{+}+T,t_{-})$, where $T=\tau -T_{p}$%
. In this case, the two-photon state is now entangled both in polarization
and in space-time, in other words, it is doubly entangled. The space-time
components $\hat{\psi}_{HH}(t_{+},t_{-})$ and $\hat{\psi}_{VV}(t_{+},t_{-})$
provide the distinguishing information for the interfering path of the
two-photon amplitudes and degrade polarization entanglement. The full
density matrix associated with the state given in Eq. \ref{eq10} can be
written as 
\begin{equation}
\rho _{ijkl}(t_{1},t_{2},t_{3},t_{4})=_{i}\left\langle j,t_{2}\right|
_{s}\left\langle i,t_{1}|\Psi \right\rangle \left\langle \Psi
|k,t_{3}\right\rangle _{s}\left| l,t_{4}\right\rangle _{i},  \label{eq11}
\end{equation}%
where $i,j,k,l=H$ or $V$. Since our chief concern is the polarization state
of the photon pair, we will calculate the effective density matrix of the
two-photon state after the two crystals that are defined in the polarization
space alone. It can be calculated by partially tracing the density matrix in
Eq. \ref{eq11} over the time-variable, and is given by 
\begin{equation}
\rho _{ijkl}=\int \int dt_{1}dt_{2}\rho _{ijkl}(t_{1},t_{2},t_{1},t_{2}).
\label{eq12}
\end{equation}%
Explicit calculation gives the effective density matrices of the form 
\begin{eqnarray}
\rho &=&\frac{1}{2}(\left| HH\right\rangle \left\langle HH\right| +\left|
VV\right\rangle \left\langle VV\right|  \nonumber \\
&&+v(T)e^{-i\phi }\left| HH\right\rangle \left\langle VV\right| +v^{\ast
}(T)e^{i\phi }\left| VV\right\rangle \left\langle HH\right| ),  \label{eq13}
\end{eqnarray}%
where we used the normalization condition of the two-photon amplitude: 
\begin{equation}
\int \int dt_{+}dt_{-}\left| \hat{\psi}_{jj}(t_{+},t_{-})\right| ^{2}=1
\label{eq14}
\end{equation}%
and notation $\rho _{ijkl}=\left\langle ji\right| \rho \left|
kl\right\rangle $ ($i,j,k,l=H$ or $V$). In Eq. \ref{eq13}, $v(T)$ is the
convolution of the two-photon amplitudes in the time domain

\begin{eqnarray}
v(T) &=&\int \int dt_{+}dt_{-}\hat{\psi}_{HH}(t_{+},t_{-})\hat{\psi}%
_{VV}^{\ast }(t_{+},t_{-})  \nonumber \\
&=&\int \int dt_{+}dt_{-}\hat{\psi}(t_{+}+T,t_{-})\hat{\psi}^{\ast
}(t_{+},t_{-}).  \label{eq15}
\end{eqnarray}%
It should be noted that $v(T)$ is, in general, a complex number. However, in
the present experiment, we always make $v(T)e^{-i\phi }$ a real number $%
\left| v(T)\right| $ by using the Bereck-type polarization compensator.
Then, the density matrix given in Eq. \ref{eq13} is formally the same as
what we observed in the experiment, and Eq. \ref{eq15} indicates that their
concurrence is determined by the convolution of the two-photon amplitudes in
the time domain.

\begin{figure}[tbp]
\epsfxsize=9.0cm
\centerline{\epsfbox{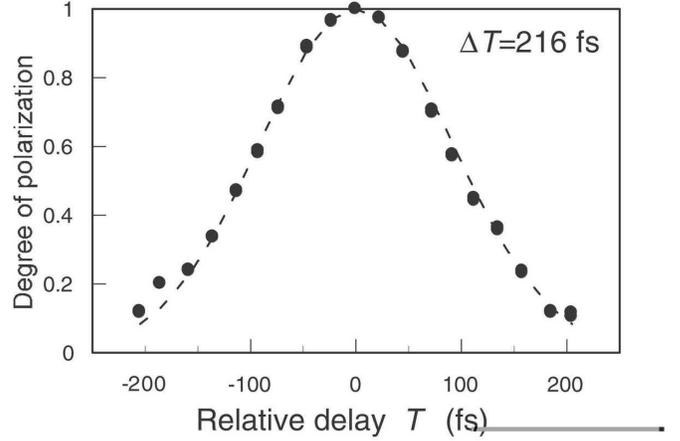}}
\caption{Autocorrelation of the pump field evaluated by measuring the degree
of polarization after the pre-compensator{\it . }They are plotted against
the relative delay introduced between two orthogonal components of the pump
field. The broken line is Gaussian fit to the data with 216-fs widths
(FWHM). }
\label{F7}
\end{figure}

\begin{figure}[tbp]
\epsfxsize=16.0cm
\centerline{\epsfbox{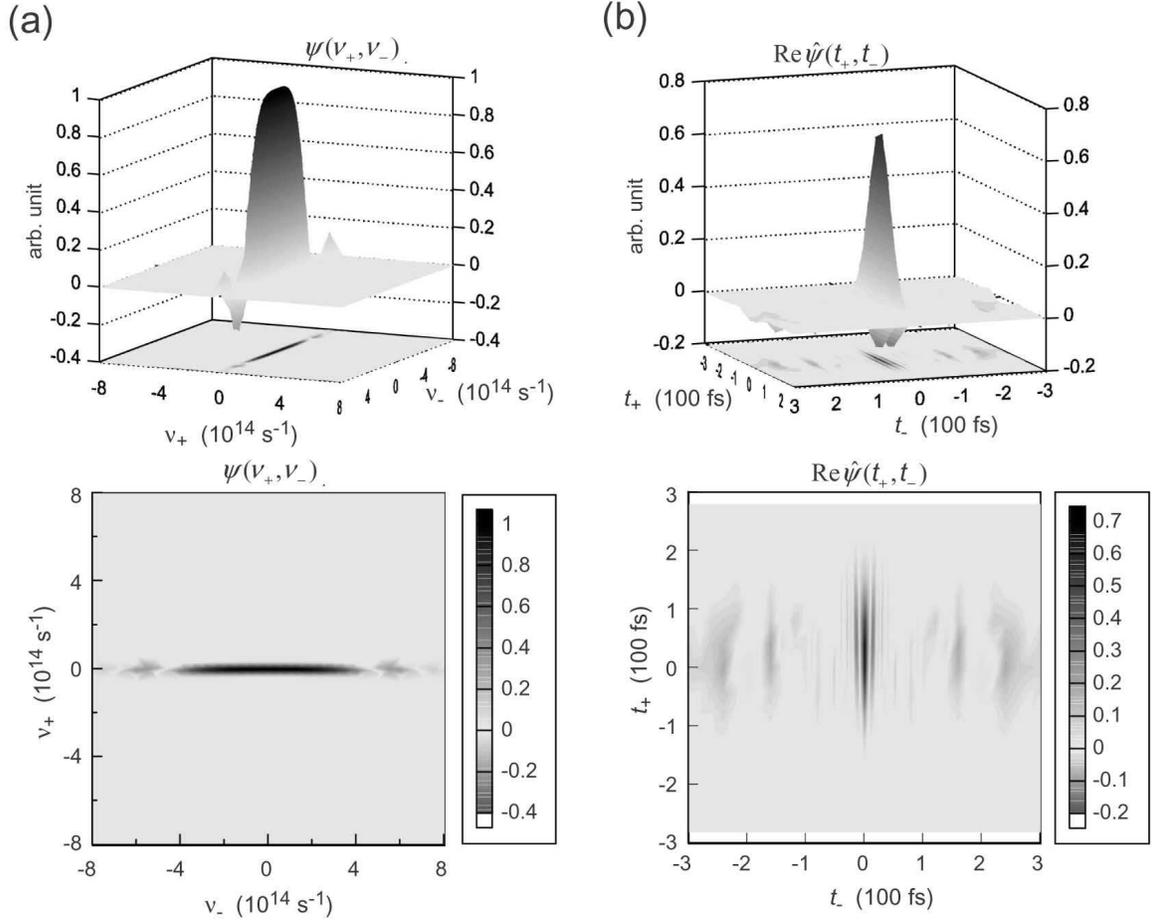}}
\caption{Calculated (unnormalized) two-photon amplitudes (a) in the
frequency domain and (b) in the time domain. Only the real part of the
amplitude is shown for the time domain. The contribution of the imaginary
part is relatively lower.}
\label{F9}
\end{figure}

\begin{figure}[tbp]
\epsfxsize=9.0cm
\centerline{\epsfbox{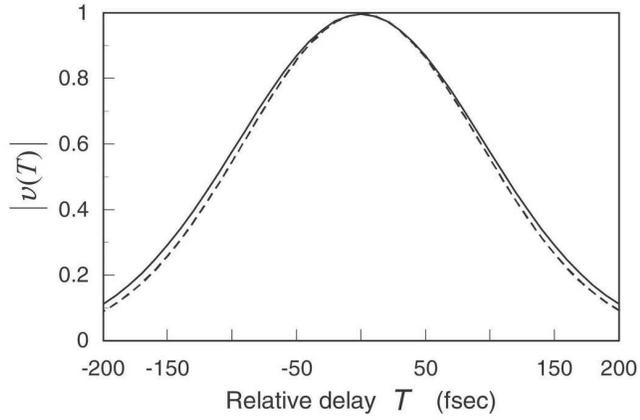}}
\caption{Calculated magnitude of the convolution $\left| v(T)\right| $. The
broken line shows the autocorrelation of the pump field.}
\label{F10}
\end{figure}

Based on the above theory, we have made several numerical calculations to
elucidate the physics behind the measured $V(T)$ curves. We calculated
two-photon amplitudes in the frequency and time domains using the
experimental parameters: $D_{+}\approx -570$ fs/mm, $D^{\prime \prime
}\approx 855$ fs$^{2}$/mm, $L=0.13$ mm, and $\tau _{p}^{FWHM}=153$ fs\cite%
{Eimerl87}. Figure \ref{F9} shows (a) the (unnormalized) two-photon
amplitude $\psi (\nu _{+},\nu _{-})$ and (b) $\hat{\psi}(t_{+},t_{-})$. We
can see that the bandwidth of $\psi (\nu _{+},\nu _{-})$ is much wider in
the $\nu _{-}$-direction than in the $\nu _{+}$-direction, while the
temporal width of $\hat{\psi}(t_{+},t_{-})$ is much wider in the $t_{+}$%
-direction than in the $t_{-}$-direction. This is due to the
Fourier-transform relations between $\psi (\nu _{+},\nu _{-})$ and $\hat{\psi%
}(t_{+},t_{-})$. We can also see that the temporal width of $\hat{\psi}%
(t_{+},t_{-})$ in the $t_{+}$-direction is $\approx \tau _{p}^{FWHM}$. Using
these results, the magnitude of the convolution $\left| v(T)\right| $ is
calculated and plotted in Fig. \ref{F10} together with the fitted curve of
autocorrelation of the pump field shown in Fig. \ref{F7}. We can see that $%
\left| v(T)\right| $ nearly agrees with the autocorrelation curve of the
pump field. This indicates that the temporal shape of $\hat{\psi}%
(t_{+},t_{-})$ is approximately Gaussian in the $t_{+}$-direction and their
widths are nearly equal to the temporal width of the pump field. This is
consistent with the fact that, although there is a small difference in the
width, the measured $V(T)$ curves nearly agree with the autocorrelation
curve of the pump field as in Figs. \ref{F5}-\ref{F7}.

Now, let us examine\ in more detail the origin of the different widths of
the measured $V(T)$ curves in Figs. \ref{F5} and \ref{F6} in which the
different bandwidth spectral filters were used. We will show that this was
due to spectral filtering effects. So far, the two-photon amplitudes after
the crystal and before the spectral filter have been discussed. They are
intrinsic ones in the sense that they are determined only by the pump field
and the crystal parameters and irrelevant to the spectral filters. In
contrast, what we actually observed in the experiment was the two-photon
state after the spectral filters. Hence, it is required to relate the state
after the filters to the state before the filters to find the spectral
filtering effects. This can be done straightforwardly. Suppose that the
linear spectral filters used for the signal and idler have complex frequency
responses expressed by $F_{s}(\nu _{s})$ and $F_{i}(\nu _{i})$,
respectively. These functions are assumed to be normalized such that 
\begin{equation}
\int d\nu _{s}\left| F_{s}(\nu _{s})\right| ^{2}=\int d\nu _{i}\left|
F_{i}(\nu _{i})\right| ^{2}=1  \label{eq16}
\end{equation}%
holds. Then the two-photon amplitude after the filters are simply given by $%
\Psi _{F}(\nu _{s},\nu _{i})=\Psi (\nu _{s},\nu _{i})F_{s}(\nu
_{s})F_{i}(\nu _{i})$. Since the two-photon amplitude in the time domain is
related to that in the frequency domain by the Fourier-transform relations,
the associated two-photon amplitude in the time domain is written as 
\begin{eqnarray}
\hat{\Psi}_{F}(t_{s},t_{i}) &=&\frac{e^{-i\Omega _{p}t_{s}}}{2\pi }\int \int
d\nu _{s}d\nu _{i}e^{-i(\nu _{s}t_{s}+\nu _{i}t_{i})}\Psi _{F}(\nu _{s},\nu
_{i})  \nonumber \\
&=&\frac{1}{\left( 2\pi \right) ^{2}}\int \int d\tau _{s}d\tau _{i}\hat{\Psi}%
(\tau _{s},\tau _{i})\hat{\Phi}(t_{s}-\tau _{s},t_{i}-\tau _{i}),  \nonumber
\\
&&  \label{eq17}
\end{eqnarray}%
where $\hat{\Phi}(t_{s},t_{i})=\hat{F}_{s}(t_{s})\hat{F}_{i}(t_{i})$ is a
normalized two-variable function that incorporates the temporal response of
the filters, and $\hat{F}_{s}(t_{s})$ and $\hat{F}_{i}(t_{i})$ are the
Fourier transform of $F_{s}(\nu _{s})$ and $F_{i}(\nu _{i})$, for example, 
\begin{equation}
\hat{F}_{s}(t_{s})=\frac{1}{\sqrt{2\pi }}\int d\nu _{s}e^{i\left( \nu
_{s}+\Omega _{p}/2\right) t_{s}}F_{s}(\nu _{s}).  \label{eq18}
\end{equation}%
Equation \ref{eq17} indicates that the two-photon amplitude after the
filters is given by the convolution of the two-photon amplitude before the
filters $\hat{\Psi}(t_{s},t_{i})$ and the apparatus function for the filters 
$\hat{\Phi}(t_{s},t_{i})$. Because of this relation, the input function $%
\hat{\Psi}(t_{s},t_{i})$ is smoothed by the filter function $\hat{\Phi}%
(t_{s},t_{i})$\ to yield broadened output function $\hat{\Psi}%
_{F}(t_{s},t_{i})$. It may be worth pointing out that Eq. \ref{eq17} is
formally analogous to the connection of phase-space probability distribution
with the Wigner functions describing system and measuring apparatus (quantum
ruler or filter)\cite{Wodkiewicz84,Knight95}. In this case, the concurrence
and visibility are determined by the convolution of the output function $%
\hat{\psi}_{F}(t_{+},t_{-})=\hat{\Psi}_{F}((t_{s}+t_{i})/2,(t_{s}-t_{i})/2)$
in contrast to Eq. \ref{eq15}, in which the input function $\hat{\psi}%
(t_{+},t_{-})$ appears. Accordingly, their widths depend on the temporal
response of the filters, and equivalently, depend on the bandwidth of the
filters. In general, the smaller the filter bandwidth is, the larger the
width of the $V(T)$ curve is, and vice versa.

Now, let us discuss quantitatively the difference in the width of the
measured $V(T)$ curves associated with the 8-nm and 40-nm filters. If we
assume the frequency response functions $F_{s}(\nu _{s})$ and $F_{i}(\nu
_{i})$ to be well approximated by the Gaussian function with FWHM widths $%
\Delta \nu $, the temporal widths of the associated response functions $\hat{%
F}_{s}(t_{s})$ and $\hat{F}_{i}(t_{i})$ is in proportion to $\Delta \nu
^{-1} $. In our case, $\Delta t_{8nm}\approx 104$ fs for the 8-nm filter and 
$\Delta t_{40nm}\approx 20.8$ fs for the 40-nm filter. On the other hand,
from the experimental results shown in Figs. \ref{F5} and \ref{F6}, those
consistent with the present theory are roughly estimated to be $\Delta
t_{8nm}^{\prime }\approx 86.0$ fs and $\Delta t_{40nm}^{\prime }\approx 17.2$
fs. Here, we used the fact that the width $\sigma $ of the convolution of
two Gaussian functions each having width $\sigma _{1}$ and $\sigma _{2}$ is
given by $1/\sigma ^{2}=1/\sigma _{1}^{2}+1/\sigma _{2}^{2}$. Therefore,
even if there remain small quantitative discrepancies, it seems reasonable
to conclude that there is considerable validity in the above simplified
theory.

\begin{figure}[tbp]
\epsfxsize=9.0cm
\centerline{\epsfbox{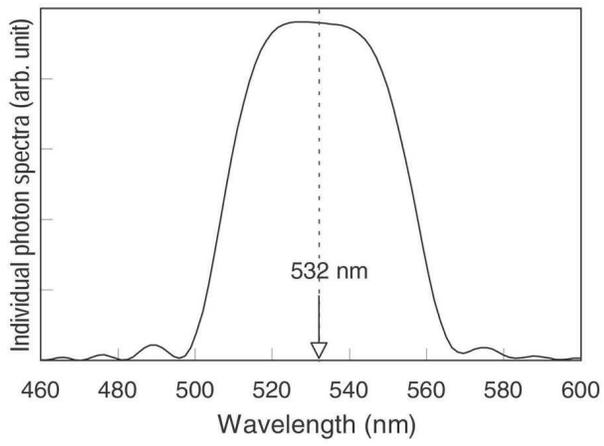}}
\caption{Calculated spectrum for the down-converted photons just after the
crystal. Slight assymetry in the spectrum with respect to the phase-matching
wavelength (=532 nm) is due to the combined effect of the group velocity
mismatch and the group velocity dispersion in the nonlinear crystal.}
\label{F11}
\end{figure}

Let us switch our attention to the spectral characteristics. For the
two-photon state given in Eq. \ref{eq1}, the joint spectral intensity is
given by $I(\nu _{s},\nu _{i})\propto \left| \Psi (\nu _{s},\nu _{i})\right|
^{2}$. Then the spectra $I_{s}(\nu _{s})$ and $I_{i}(\nu _{i})$ of the
signal and the idler can be obtained by tracing $I(\nu _{s},\nu _{i})$ over
the unobserved variable, for example, 
\begin{equation}
I_{s}(\nu _{s})\propto \int d\nu _{i}I(\nu _{s},\nu _{i}).  \label{eq19}
\end{equation}%
Figure \ref{F11} shows the calculated individual spectra for the
down-converted photons just after the crystal. This figure indicates that
the FWHM bandwidth of the down-converted photons is $\approx 40$ nm. Thus,
in our experiment using the 8-nm filters in front of the detectors, what we
have observed is only a fraction of the photons having the wavelength $%
532\pm 4$ nm that can reach the detectors. On the contrary, we can safely
say that almost all the photons produced in the SPDC have been observed in
our experiment using the 40-nm filters. In other words, the coincidence
count rate ($\approx 2700$ cps) obtained in the experiment using the 40-nm
filters had already reached the upper limit in our configuration. However,
since no optimization was made for the transversal spatial mode of the
down-converted photons in this experiment, we believe that further
improvement in the coincidence count rate may be possible by optimizing the
experimental configuration\cite{Kurtsiefer01,Monken98}.

While our simplified model seems satisfactory in essence, there still
remains a problem which needs to be solved. Within our model, the
autocorrelation curve of the pump field should have the smallest width,
whereas this is not the case. Although we have not arrived at a conclusion
yet, there might be things neglected in this consideration that should
actually have been taken into consideration. For example, we neglected the
effect higher than the second-order optical nonlinearity in the crystal.
There actually might occur higher order effects. In particular, it is likely
that self-phase modulation (SPM) due to third-order optical nonlinearity
occurs when the crystals are pumped by ultrashort optical pulses having
large peak intensities. When the SPM is taken into account, the two-photon
state after the crystal is chirped, and their coherence length might be
shortened, and the resultant width of the measured $V(T)$ curves might be
reduced. To obtain a more satisfactory explanation, the effect of SPM in the
crystals might be incorporated into the theory.

\section{CONCLUSION}

\label{SEC4}

In conclusion, we have generated pulsed polarization-entangled photon pairs
by femtosecond-pulse-pumped SPDC in a cascade of two type-I crystals. It was
found that highly entangled photon pairs were successfully obtained by using
giving an appropriate temporal delay between the orthogonal polarization
components for the pump, without the need of spectral post-selection using
narrow-bandwidth filters. Theoretical analysis showed that entanglement
depends on the convolution of the space-time components of the interfering
two-photon amplitudes. The experimental results obtained were well explained
if the dispersion and birefringence effects in the two-crystals are taken
into account. Our analysis also revealed the effect of the spectral
filtering on the magnitude of the entanglement when the interfering
two-photon amplitudes has a different space-time dependence.

This work was supported by CREST of JST (Japan Science and Technology
Corporation).


\onecolumn

\twocolumn

\onecolumn

\twocolumn

\end{document}